\documentclass[letter]{ieice}
\usepackage[dvips]{graphicx}
\usepackage[fleqn]{amsmath}
\usepackage{times}
\usepackage{url}
\usepackage{setspace}
\usepackage{ifthen,amssymb}
\usepackage{amsmath}
\usepackage{cases}
\usepackage{listings}
 \usepackage{caption}
\usepackage{float}
\usepackage{color}
\usepackage{multirow}
\usepackage{booktabs}
\usepackage{amsmath}
\usepackage{mathtools}
 \usepackage{listings}
\usepackage[utf8]{inputenc}
\usepackage[T1]{fontenc}
\usepackage{cite}
\usepackage{amsmath,amssymb,amsfonts}
\usepackage{algorithmic}
\usepackage{graphicx}
\usepackage{textcomp}
\usepackage[font=small,skip=0pt]{caption}

\newcommand{\code}[1]{\texttt{#1}}
\usepackage{xcolor}

\newboolean{showcomments}
\setboolean{showcomments}{true}
\ifthenelse{\boolean{showcomments}}
  {\newcommand{\nb}[2]{\fbox{\bfseries\sffamily\scriptsize#1}{\sf\small$\blacktriangleright$\textit{\scriptsize#2}$\blacktriangleleft$}}}
  {\newcommand{\nb}[2]{}}

\field{}
\title{Software Engineering Data Analytics: A Framework Based on a Multi-layered Abstraction Mechanism}
\authorlist{%
 \authorentry{Chaman Wijesiriwardana}{m}{label1}
 \authorentry{Prasad Wimalaratne}{n}{label1}
}
\affiliate[label1]{The author is with the University of Colombo School of Computing, Reid Avenue, Colombo 
}

\received{18}{3}{1}
\revised{18}{3}{1}

\begin{document}
\maketitle
\begin{summary}
This paper presents a concept of a domain-specific framework for software analytics by enabling querying, modeling, and integration of heterogeneous software repositories. The framework adheres to a multi-layered abstraction mechanism that consists of domain-specific operators. We showcased the potential of this approach by employing a case study.
\end{summary}

\section{Introduction}
\label{sec:intro}

With the increase in standard processes and productivity tool support, both the number of artifacts created in the software development process and the amount of available data has significantly grown.
These vast amounts of data contain valuable information that could help supporting software developers, but it is impossible to examine everything manually.
As a result, recent years have witnessed extensive studies on mining software repositories (MSR) to discover various types of valuable information about software projects\cite{hemmati2013msr},\cite{roehm2012professional}.
However, even with the existing MSR tools, software developers often suffer from a scarcity of techniques necessary for flexible querying and integration of custom data stored in such repositories towards accomplishing complex analysis tasks\cite{hassan2008road},\cite{sakamoto2013visualizing}. Therefore, software researchers have to spend considerable time on integrating and synthesizing the information stored in repositories to perform software evolution analysis tasks. This has identified as an essential future direction in mining software repositories\cite{gonzalezchallenges}.

Existing frameworks that support MSR based on either generic query languages such as SQL or domain-specific languages that specifically designed for MSR. Approaches such as Gitana\cite{cosentino2015gitana}, AlitheiaCore\cite{gousios2014conducting} and MetricMiner\cite{sokol2013metricminer} are based on the standard SQL syntax. These attempts are not explicitly facilitating MSR specific complex tasks such as \textit{finding the number of critical issues resolved by the most frequent committer in a project} or \textit{finding the bug introducing changes between two successful builds}.  Boa \cite{dyer2013boa} and QWALKEKO  \cite{stevens2014querying} are domain specific languages specifically designed for MSR. However, these approaches are built for querying particular data sources. Beyond that, such domain-specific languages are lacking the rich features of well-established generic query languages and not easily extended to facilitate heterogeneous data sources. 

As a solution, we introduce a conceptual foundation of a domain-specific framework to support flexible querying and analysis of software engineering data. The framework adheres to an extensible multi-layered abstraction mechanism that consists of a collection of domain-specific operators. The domain-specific operators are built on top of the operators derived from relational algebra. The framework helps both researchers and practitioners in effectively exploring various kinds of information stored in both traditional and non-traditional repositories. 

\section{Approach}
\label{sec:query_language}

Software data analytics typically consists of three phases:
It starts with collecting required data, e.g., by extracting data available on software repositories. Then the data needs to be preprocessed using various operations such as data cleaning or data filtering. Finally, the actual analysis task required for accomplishing the analysis goals can perform. To facilitate this workflow, this paper introduces a collection of software analytics specific operators, which could be meaningfully composed to build up the logic to accomplish analysis tasks.

\subsection{Operator Stack}
\begin{figure}[ht]
    \centering
    \includegraphics[width=0.9\columnwidth]{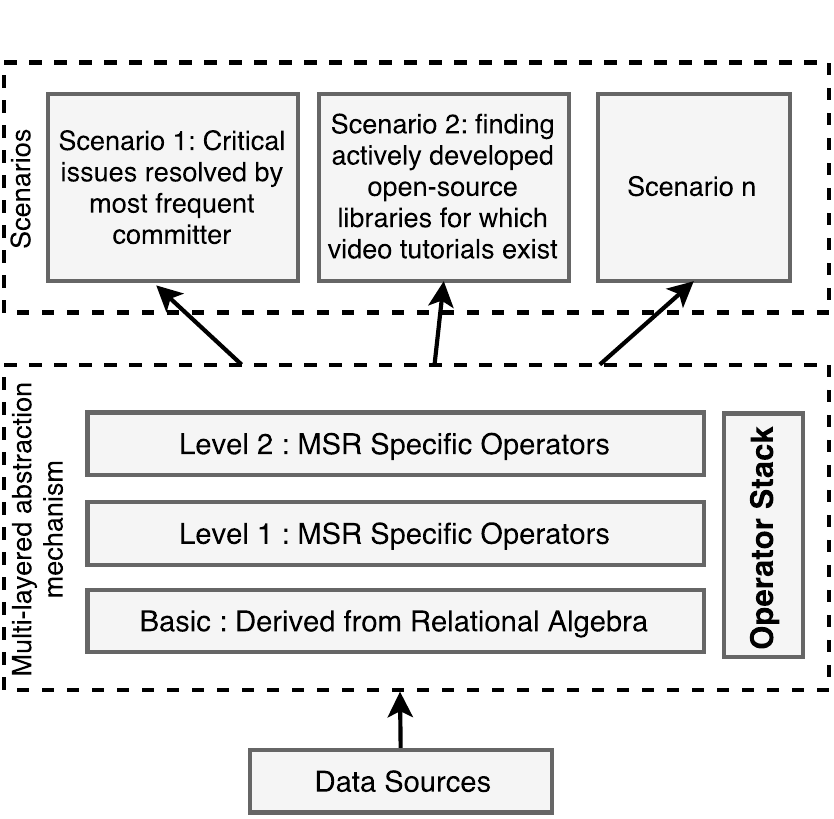}
    \caption{Operator Stack and the Examples of Supported Scenarios}
    \label{fig:stack}
\end{figure}

As depicted in Figure \ref{fig:stack}, the operator stack consists of several layers, which could be extensible with new layers and a collection of corresponding operators. In the following, we briefly describe each layer and the rationale for separating it into multiple layers.   

\textbf{Basic operators:} These operators directly derived from relational algebra. Relational algebra is a procedural query language, which operates on input relations. It consists with a set of fundamental operators such as \code{select}, \code{project}, \code{union} and \code{cartesian product}. Though several relational algebra theorems do not strictly hold in query languages such as SQL and LINQ, they are the native implementations of the underline concept of relational algebra. Therefore, we borrowed some ideas from such languages to identify basic operators supported by relational algebra. In this paper, operators such as \code{filter}, \code{select}, \code{join}, \code{sort}, \code{count}, etc has been categorized as basic level operators. Such operators are useful in generating queries to perform simple tasks such as \textit{counting the number of commits of particular developer} or \textit{finding the successful builds within a particular period}. 

\textbf{Level 1 operators:} Level 1 operators derived by combining basic operators without further processing to ease MSR specific tasks. Consider an analysis task of \textit{finding the most frequent committer in a project}. It requires extracting commit details from a VCS, filtering the required fields, sorting them according to the number of commits, and selecting the committer with the highest number of commits. The whole process could be accomplished by using a single \code{FindMax} operator, which is made up of three basic level operators: \code{filter}, \code{sort} and \code{select}. Therefore, \code{FindMax} is categorized as a Level 1 operator in the proposed operator stack. It can be used to find, for instance, \textit{finding the developer who fixed the most number of bugs}.

\textbf{Level 2 operators:} Software analysis tasks, by nature, are more complex than the examples described in the previous section. It may include complex tasks such as \code{issue-revision linking}, \code{change distilling} or \code{detecting code smells}. As a result, an operator such as \code{FindMax} could be used to perform a sub-task in a more complex analysis task. Therefore, we emphasize the critical need for more powerful operators as a single unit that is made up of a combination of basic and Level 1 operators to perform complex analytics experiments.

\subsection{Repository of operators}

\begin{figure}[ht]
    \centering
    \includegraphics[width=0.9\columnwidth]{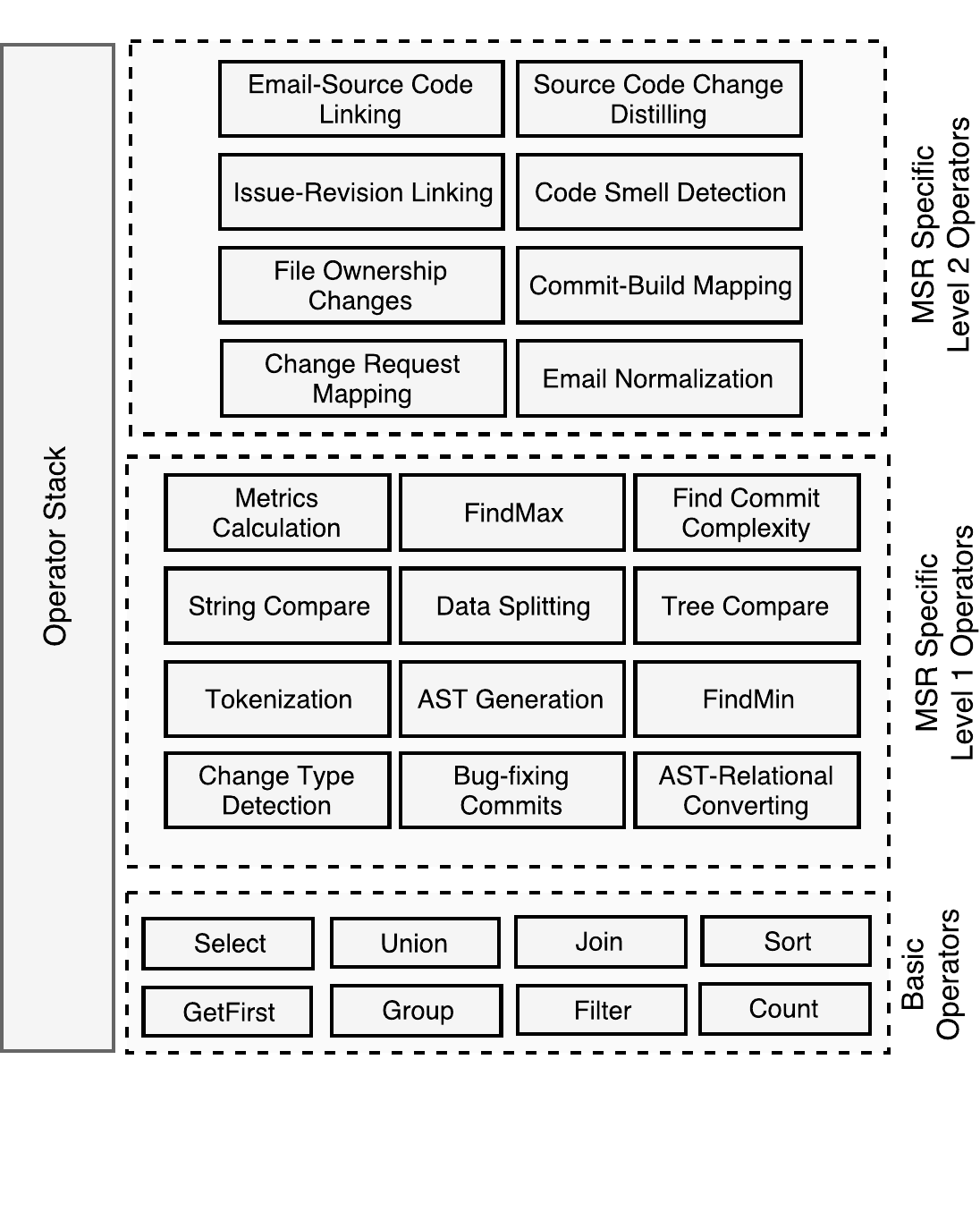}
    \vspace*{-10mm}
    \caption{Collection of Operators Derived from the Literature}
    \label{fig:repo}
\end{figure}

One of the main contributions of this paper is to introduce a collection of frequently used operators that are used to perform software analytics experiments. Initially, a literature review is conducted on the papers published on the International Conference on Mining Software Repositories\footnote{www.msrconf.org/} for the last three years (2015-2017). We carefully investigated the methodology section in each paper to identify the main analytics tasks carried out during the experiments. Our study was not limited to the baseline papers on the mentioned conference and the duration. We further traced back to find related work published outside the selected papers to identify the frequent nature of the operators. Figure \ref{fig:repo} presents the identified operators from the literature review categorized into \textit{basic}, \textit{level 1}, and \textit{level 2}. However, the operator repository would experience a continuous evolution with the time.    

\vspace*{-5mm}
\subsection{Prototype Implementation}
As a proof-of-concept, the operators are built using java languages. As of now, users can accomplish a software analysis task by writing a simple program by utilizing the available operators in the operator library. 

\section{Case Study Based Evaluation}
\label{sec:evaluation}

A case study is employed to demonstrate how to build the logic to perform a software analysis task by utilizing the operators from the operator stack. The following analysis task showcases a possible interlinking of two software repositories by utilizing both the basic level and Level 1 operators from the operator stack. 

\setlength{\parindent}{0cm}
\textbf{Analysis Task:} \textit{Finding critical issues resolved by most frequent committer in a project.} 

Measuring the performance of the developers is a non-trivial task for project managers, especially when the team size and project scope is large. Due to its subjective nature, finding an accurate, quantifiable metrics is a holy grail for managers. Simplistically, total lines of codes, number of commits, number of bugs fixed, or a meaningful combination of them could yield useful insights of performance.

 \textbf{Background:} Software artifacts produced during the development of a software project are disconnected from each other. Establishing traceability links across artifacts is a key challenge in software maintenance \cite{witte2007empowering}.  To address this, gold rush for querying the interrelationship data across autonomous and heterogeneous software repositories has witnessed in several recent studies.

\begin{figure}[ht]
  \centering
\includegraphics[width=80mm]{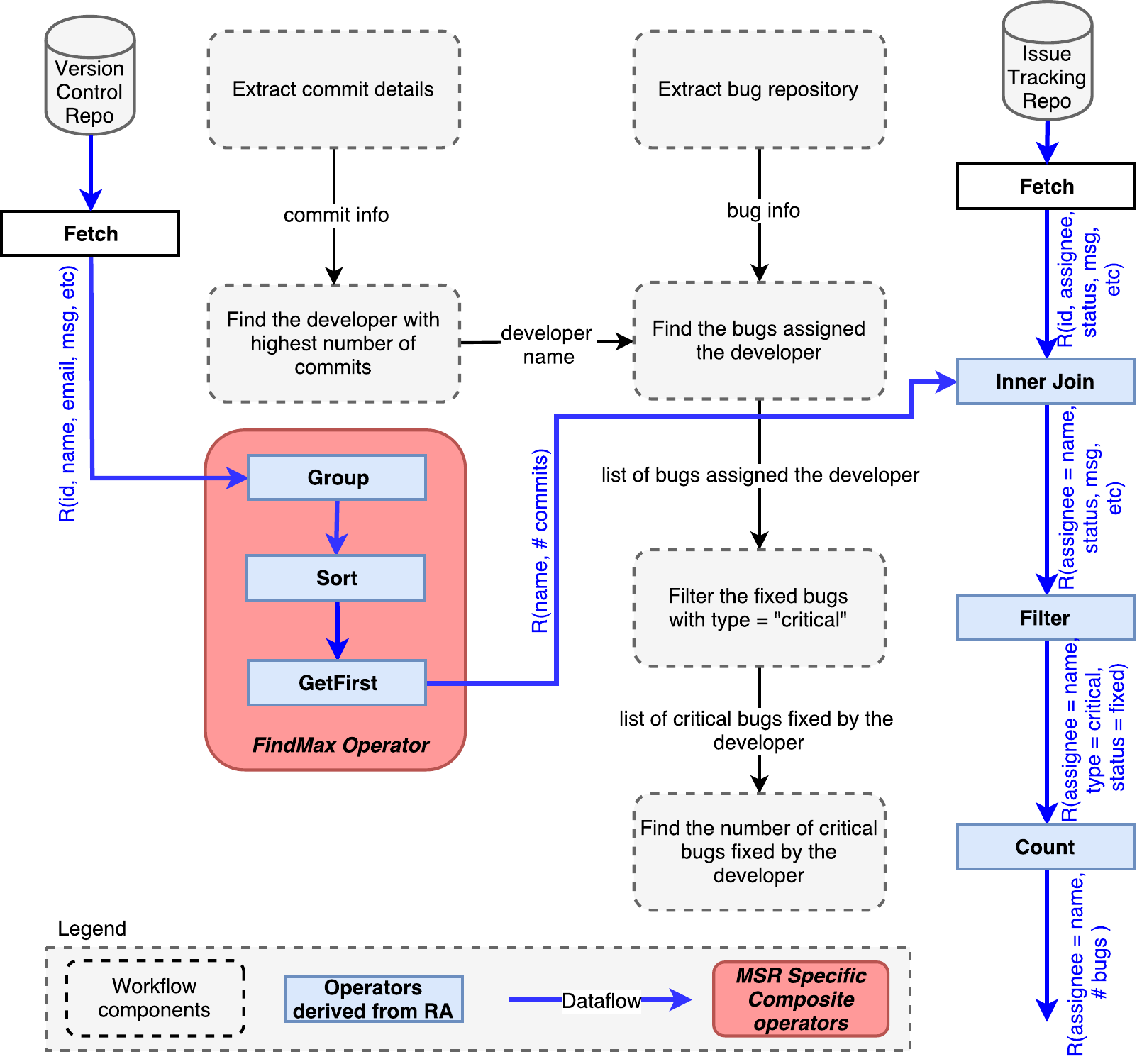}
       \caption{Finding the number of critical issues resolved by the most frequent committer in a project}
   \label{fig:fr_committer}
\end{figure}

\textbf{Implementation using operators:} Figure \ref{fig:fr_committer} presents how to utilize the identified operators to perform the above analysis task. It demonstrates how the data is integrated from both version control and bug tracking repositories to find how many critical bugs have fixed by the most frequent committer.    

\vspace*{-2mm}

We further tested the above scenario with four open source projects by using the prototype implementation. The prototype allows users to utilize the operators to perform the tasks directly. Therefore, it provides a great level of convenience to the users.  Summary of the experimental results present in Table \ref{tab:summary}. The names of the developers keep anonymous on the table, however, can be exposed to the reviewers.   

\begin{table}[h]
\small
\caption{Summary of the Experimental Results}
\begin{tabular}{lp{1cm}lp{1cm}lp{1cm}l}
\toprule
Apache Project & No: of Commits                                                & Frequent Developer                    & No: of bug fixes          \\ \midrule 
Gora  & 1053 & Developer A     & 52   \\
Commons-lang  & 5171 &  Developer B     & 4   \\
IO  & 2091 & Developer C  & 0 \\
Winx  & 1312       & Developer D & 2 \\
\midrule 
\end{tabular}
\label{tab:summary}
\end{table}

\section{Conclusion}
\label{sec:conclusion}

This paper presents a conceptual foundation of a domain-specific framework based on a multi-layered abstraction mechanism to support flexible querying and analysis of software engineering data. The framework supports querying, modeling, and integration of heterogeneous and autonomous software repositories. Main building blocks of the multi-layered abstraction mechanism are represented regarding a collection of basic and MSR specific operators. As future work, we plan to enrich the operator stack with more operators derived from the MSR research. Then the framework will be evaluated with more case studies by incorporating a diverse set of analysis tasks.

\begin{flushleft}
\textbf{Acknowledgements}
\end{flushleft}
The authors of this paper gratefully acknowledge the financial support provided by the National Research Council (Grant No: NRC 15-74). We thankfully acknowledge the insights and expertise provided by the colleagues at SEAL Lab at the University
of Zurich.

\bibliographystyle{ieicetr}
\bibliography{references}


\end{document}